# Emergent Geometric Frustration of Artificial Magnetic Skyrmion Crystals

Fusheng Ma,[1] C. Reichhardt,[2] Weiliang Gan,[1] C. J. Olson Reichhardt,[2] and Wen Siang Lew,[1,*]

[1]School of Physical and Mathematical Sciences, Nanyang Technological University, 21 Nanyang Link, Singapore 637371.

[2]Theoretical Division and Center for Nonlinear Studies, Los Alamos National Laboratory, Los Alamos, New Mexico 87545, USA.

*Corresponding author: wensiang@ntu.edu.sg

Magnetic skyrmions have been receiving growing attention as potential information storage and magnetic logic devices since an increasing number of materials have been identified that support skyrmion phases. Explorations of artificial frustrated systems have led to new insights into controlling and engineering new emergent frustration phenomena in frustrated and disordered systems. Here, we propose a skyrmion spin ice, giving a unifying framework for the study of geometric frustration of skyrmion crystals in a non-frustrated artificial geometrical lattice as a consequence of the structural confinement of skyrmions in magnetic potential wells. The emergent ice rules from the geometrically frustrated skyrmion crystals highlight a novel phenomenon in this skyrmion system: emergent geometrical frustration. We demonstrate how skyrmion crystal topology transitions between a non-frustrated periodic configuration and a frustrated ice-like ordering can also be realized reversibly. The proposed artificial frustrated skyrmion systems can be annealed into different ice phases with an applied current induced spin-transfer torque, including a long range ordered ice rule obeying ground state, as-relaxed random state, biased state and monopole state. The spin-torque reconfigurability of the artificial skyrmion ice states, difficult to achieve in other artificial spin ice systems, is compatible with standard spintronic device fabrication technology, which makes the semiconductor industrial integration straightforward.

# I. INTRODUCTION

Recently, a topological particle-like nanometre-sized spin texture, the magnetic skyrmion, has gained attention. These nontrivial spin textures were theoretically predicted to exist in certain non-centrosymmetric magnetic materials [1] with Dzyaloshinskii-Moriya (DM) interactions [2,3] and were subsequently experimentally identified in the chiral magnet MnSi by neutron scattering [4], Lorentz transmission electron microscopy (LTEM) [5,6], topological Hall effect [7–9], and Dynamic Cantilever Magnetometry [10]. Interface-induced skyrmions have been experimentally observed in perfectly ordered, atomically thin layers by spin-resolved scanning tunneling microscopy (STM) [11,12] and time-resolved pump-probe X-ray holography [13]. Local currents from a STM tip can be used to write and delete individual isolated skyrmions [12]. The generation and movement of interface-induced skyrmionic bubbles at room temperature has also been realized using a geometric structure to "blow" the bubbles into a magnetic layer grown epitaxially on a material with large spin-orbit coupling [14]. Very recently, the experimental observation of nanoscale magnetic skyrmions (< 100 nm) has been achieved in ultrathin metallic films and multilayers at room temperature or above and in the absence of an external magnetic field: Pt/CoB/Pt [13], Pt/Co/Ta [15], Pt/CoFeB/MgO [15], Ir/Co/Pt [16], Pt/Co/MgO [17], and Ir/Fe/Co/Pt [18]. Furthermore, the presence and nucleation of individual/isolated skyrmion in confined magnetic nanostructures as well as the evolution of the skyrmion size has also been demonstrated in circular dots [13,15–18] and nanowires [15,16]. These achievements are highly promising for future skyrmion-based devices. There has been tremendous interest in magnetic skyrmions due to their unusual spintronic properties, such as their topological stability which protects them from being hindered by defects [19–21] and their current-driven motion with an ultralow depinning threshold current [21–23]. In view of these distinct features, magnetic skyrmions are promising candidates for information carriers in a range of applications in spintronics such as race-track memory devices [20,21], skyrmion logic devices [24], or skyrmion magnonic devices [25], so understanding how to systematically control skyrmion configurations and dynamics with nanopatterns has broad relevance. Although numerous efforts have been devoted to manipulating skyrmion motion, there is little work on how to precisely control their position and confinement.

In contrast to the self-assembled triangular skyrmion lattice stabilized by DM interaction in some non-centrosymmetric magnetic materials [1,4,5,11,19,26], new approaches are suggested for creating and stabilizing two-dimensional artificial lattices of magnetic skyrmions by periodic modulation of either the geometrical [27–33] or the material properties of the magnetic thin films [34]. In this work, artificial skyrmion crystals with either square or honeycomb lattices have been created in periodically nanopatterned magnetic thin films. The structural confinement of skyrmions in magnetic potential wells causes geometric frustration of skyrmion crystals in a non-frustrated artificial geometrical lattice.

Geometric frustration arises in a variety of natural systems, with some of the archetypes including proton ordering in water ice [35] and magnetic ordering in pyrochlore crystals [36–38]. The latter are referred to as spin ice [39,40]. The local elementary excitations of spin ices have generated considerable interest in recent years because that they can be parameterized as emergent magnetic monopoles [41,42]. Their three-dimensional atomic lattices situate rare-earth ions with large magnetic moments on the corners of tetrahedra. Although the naturally occurring spin ices exhibit interesting types of topological monopole defects and "ice-rule" states [39,40], the frustrated behaviors occur only at very low temperatures and the individual spin ordering or defects cannot be directly visualized on the atomic scale size. Recent advances in nanofabrication technology have permitted the creation of artificial ice systems [43–62] that mimic the behavior of geometrically frustrated atomic spin ices at much larger length scales and higher temperatures, where direct visualization of the microscopic effective spin configurations under controlled conditions is possible.

The most studied artificial spin ice systems are created using arrays of nanomagnets with a bistable single-domain magnetization [43,44,46–57,63–67]. Artificial nanomagnet spin ice systems have been realized experimentally for square [43,50,51,56,66], and honeycomb [49,52,68] lattices, each having different analogous features to the naturally occurring rare-earth pyrochlore lattice [37]. In these ice systems, each nanomagnet plays the role of an effective macrospin, and the spin direction is defined to point in the direction of the magnetic moment. The vertex state of the system can be directly imaged by locally probing the magnetic moment of a single constituent nanomagnet via magnetic microscopy, such as with magnetic force microscopy [53], LTEM [69], or photoemission electron microscopy [68]. In square nanomagnet ices [43,50,51,56,66], however, the ice-rule-obeying states with "two-spins-in/two-spins-out" order have been only partially visualized. The completely ordered ice-rule-obeying ground state (GS) with two ''in'' spins on opposite sides of the vertex has not been observed experimentally. The observed disordered states, composed of a mixture of ice-rule-obeying and ice-rule-breaking vertices [43], may arise due to the relatively weak magnetic interactions between the nanomagnets and the relatively high energy barrier between the two possible nanomagnet states, as well as from quenched disorder effects in the nanomagnet array [44]. Recently, it was demonstrated that the GS could be approached more closely by using certain dynamical annealing protocols [43,51,70]. Alternative realizations of the artificial spin ice system can be created using colloids in arrays of elongated optical traps [58,59] or vortices in superconducting films with arrays of pinning defect sites [60–62]. In these two particle analogues, the effective spin vector is defined according to the position of a colloid or vortex in a double-well potential well. It has been shown numerically that both the colloid and vortex ice systems can produce not only the same spin ice rules observed for nanomagnet ices, but also the ice-rule-obeying ground states not yet observed in nanomagnet ices [58,60]. A colloidal version of an artificial spin ice system has been realized using interacting paramagnetic colloids [71,72]. Artificial vortex ice states have been experimentally realized in type II superconducting MoGe thin films (< 5 K) [61] and YBaCuO thin films (< 50 K) [62] with nanoscale spatial pinning arrays. It is, however, challenging to observe the vortex dynamics directly; in addition, the vortex artificial ices exist only at very low temperatures.

Here, we show that magnetic skyrmions in magnetic thin films with perpendicular anisotropy and interfacial DM interaction can be artificially arranged with appropriately designed nanostructured arrays of artificial potential wells. The phase behavior of skyrmions assembled in a square/honeycomb lattice of ellipse shaped magnetic potential wells is investigated. Either a non-frustrated or frustrated skyrmion crystal can be realized depending on the position of skyrmion located in each potential well. An external spin-polarized current is used to anneal the artificial skyrmion ice system into different ice phases, including a long range ordered ground state, as-relaxed random state, biased state and monopole state. Due to the nature of skyrmions, this system can exhibit a number of properties such as reversible structural transitions as a function of field which would be difficult to create in other artificial spin ice systems.

## II. MODEL AND METHOD

A schematic of the proposed skyrmion crystal system is shown in Fig. 1(a) which illustrates a regular square array of elliptical blind holes with lattice constant 120 nm. The blind holes are elliptical in shape with a major axis of 90 nm and a minor axis of 30 nm. The magnetic film thickness is 5 nm, and the depth and bottom thickness of the blind holes is 4 nm and 1 nm, respectively. By patterning a magnetic thin film with a square array of blind holes, a periodic array of potential wells is created in the continuous magnetic layer under the blind holes. Figure 1(b) shows the periodic potential wells indicated by the perpendicular component of the magnetic stray field in the bottom continuous layer.

### A. Creation of artificial skyrmion crystal

A magnetic skyrmion is a nontrivial spin texture of the normalized classical spin $\mathbf{m} = (m_x, m_y, m_z)$ with a whirling configuration [1,4,5,11,19,26] characterized by a topological integer winding number. Figure 1(c) displays a top view of the magnetization configuration of a single skyrmion and its corresponding topological density distribution, which has a radial symmetry distribution. The spin at the outer periphery is up, that is, (0, 0, 1), whereas the spin direction at the skyrmion center is down, that is, (0, 0, -1). The complexity of the skyrmion magnetization configuration is characterized by a topological integer skyrmion number (*i.e.* topological charge), which is defined by integrating a topological density $n$ over the plane [4,11,26,73]:

$$N = \frac{1}{4\pi} \iint n \, dx dy, \text{ with } n = -\mathbf{m} \cdot \frac{\partial \mathbf{m}}{\partial x} \times \frac{\partial \mathbf{m}}{\partial y} \qquad (1)$$

where $\mathbf{m} = \mathbf{M}(\mathbf{r})/|\mathbf{M}(\mathbf{r})|$ is the unit vector of the local magnetization indicating the orientation of the magnetization. The skyrmion number, $N = 1$, defined over a sufficiently large area, is topologically invariant against a smooth deformation of the spin texture. Thus, a skyrmion is topologically protected from destruction or splitting. In the center of the skyrmion, the value of $|n|$ is maximized. Moving from the center to the periphery, $|n|$ initially decreases and then increases to a local maximum near the skyrmion radius before finally gradually decreasing to its lowest value at the sample boundary. The topological stability of skyrmions makes them unique among different topological objects: vortex, meron, and bubble. The topological density of both vortex and meron is concentrated on the core. The topological density of a bubble is located at the domain wall between two magnetic domains, whereas that of a skyrmion is distributed in the whole object. The topological density with a radial symmetry has the largest value near the center of the skyrmion, and the smallest at the boundary. The presence and nucleation of individual/isolated skyrmions in confined magnetic nanostructures as well as the evolution of the skyrmion size has also been demonstrated in circular dots [13,15–18,74] and nanowires [15,16].

The skyrmions can be easily generated in the holes simply by subjecting the patterned film to a uniform external magnetic field [32]. As the coercivity of the perpendicular magnetic films can be significantly less than anisotropy, the magnetization reversal can start with the appearance of the reversed nucleus in the "weak" place or the bottom of the blind holes with the magnetization inside and outside the blind holes along the negative and positive out of plane direction as shown in Fig. 1(a). The system is then relaxed to a stationary state with skyrmions generated and stabilized in the holes. The stray field in the hole is aligned with the skyrmion magnetization direction and acts to confine the skyrmion in the hole. The resulting skyrmion crystal remains stable over a wide field range. The size of the skyrmions is determined by the magnetic properties of the thin film as well as the strength of the magnetic field. When the size of the skyrmions is comparable with or even larger than the minor axis of the elliptical blind holes, the skyrmions are stabilized at the centers of the holes as shown in Fig. 1(d). However, when the skyrmions are smaller than the minor axis of the elliptical blind holes, the two ends of each hole are the energetically favored places for the skyrmion to sit due to the symmetrical double well shape of the blind holes. Hence, a single skyrmion trapped in a hole has two possible lowest energy states and sits at either end of the elliptical hole depending on the interactions with nearby skyrmions as shown in Fig. 1(e).

Using an approach proposed by Thiele [75], the equation for the drift velocity of the skyrmion can be obtained by mapping the LLG equation onto the translational mode in the continuum limit, while assuming the rigidity of the spin textures during the skyrmion motion [23,76–78]. Since skyrmions are particle-like objects, their dynamical properties can be captured using a modified Theile's equation [75] that includes the current induced drag force and Magnus force, the repulsive skyrmion-skyrmion interactions, and the pinning force [23,75,77–82]:

$$\bm{G} \times (\bm{v}_s - \bm{v}_d) + \bm{D}(\beta \bm{v}_s - \alpha \bm{v}_d) = \bm{F}_i^{ss} + \bm{F}_i^{sp} \qquad (2)$$

where $\bm{v}_d$ is the drift velocity of the skyrmion while $\bm{v}_s$ is the velocity of the conduction electrons. The first term in the left hand side of equation (2) is the Magnus force with $\bm{G}$ as the gyromagnetic coupling vector. The Magnus term produces a force that rotates the velocity toward the direction perpendicular to the net external forces. The second term is the dissipative force with $\bm{D}$ as the dissipative force tensor. $\beta$ is the non-adiabatic constant of the STT as expressed by Thiaville [83]. The skyrmion-skyrmion interaction force is $\bm{F}_i^{ss} = \sum_{j=1}^{N_s} \bm{r}_{ij} B R_{ij}$, where $R_{ij} = |\bm{r}_i - \bm{r}_j|$, $\bm{r}_{ij} = (\bm{r}_i - \bm{r}_j)/R_{ij}$, and $B$ is the modified Bessel function. This repulsive interaction falls off exponentially for large $R_{ij}$. The pinning force $\bm{F}_i^{sp}$ arises from periodically arranged blind holes with a maximum pinning force of $F_p$. The driving force from an externally applied current interacting with the emergent magnetic flux carried by the skyrmions [23] is slowly increased in magnitude to avoid any transient effects.

### B. Micromagnetic simulations

The micromagnetic simulations were performed with MuMax3 [84], which incorporates the Dzyaloshinskii-Moriya interaction [85–87] and the spin-polarized current induced spin-transfer torque (STT) [83,88] as described by the modified Landau-Lifshitz-Gilbert equation. Material parameters of the nanotracks used in the simulations are those of cobalt (Co) on a platinum (Pt) substrate inducing DMI as follows [20]: the saturation magnetization $M_s = 5.8 \times 10^5$ A/m, the exchange stiffness $A = 1.5 \times 10^{-11}$ J/m, the perpendicular magnetic anisotropy $K = 0.8$ MJ/m$^3$, the DMI constant $D = 3$ mJ/m$^2$, the damping constant $\alpha = 0.1$, the nonadiabatic spin-transfer parameter $\beta = 0.3$, the gyromagnetic ratio $\gamma = 2.211 \times 10^5$ m/As, and the polarization rate of the current $P = 0.7$. The cell size used in the simulation is $2 \times 2 \times 1$ nm$^3$, which is well below the characteristic domain wall length. The calculations are started from a uniformly magnetized film with the magnetization inside and outside the blind holes along the negative and positive out of plane direction, respectively. The system then relaxes to a stationary state with skyrmions generated and stabilized in the holes. The size of the skyrmions is determined by the magnetic properties of the thin film and is smaller than the minor axis of the elliptical blind holes. Due to the symmetrical double well shape of the blind holes, the two ends of each hole are the energetically favored places for the skyrmion to sit. Hence, the single skyrmion in each hole will be located at one end of the hole after relaxation.

### C. Phase diagram of skyrmion crystal

The skyrmion size can be manipulated by applying a magnetic field $H_z$ opposite or parallel to the core of the skyrmion as shown in Fig. 2(a). The skyrmion shrinks/expands if $H_z$ is directed opposite/parallel to its core magnetization. Hence, the phase diagram of the magnetization configuration in the magnetic thin film with an array of ellipse-shaped blind holes can be divided into ferromagnetic (FM) and skyrmion crystal (SC) phases. With large positive $H_z$, the skyrmions can be annihilated inside the holes to reach the positive FM phase, while with large negative $H_z$, the skyrmions can be expanded out of the holes to reach the negative FM phase. An SC phases exists when skyrmions are present in the holes. In the SC phase as shown in Fig. 2(b), the skyrmions are located at the center of the elliptical holes and form a square skyrmion crystal.

However, as shown in Fig. 2(c) with zero field, the skyrmions shrink and shift from the center of the blind holes to the ends, which is the lowest energy position for small skyrmions. The rearrangement of skyrmion location in the square lattice of elliptical holes indicates an emergence of geometrical frustration in the square skyrmion crystal. Thus, the square skyrmion crystal undergoes a solid-solid transformation and becomes a square skyrmion ice subject to geometric frustration. The magnetic switching between a non-frustrated skyrmion crystal and a frustrated ice-like skyrmion crystal is also visible in the topological density distribution of the local skyrmions shown in Figs. 2(b) and 2(c). This ability to transform between non-frustrated and frustrated skyrmion crystals cannot be achieved in nanomagnetic systems as the nanomagnet array layout is fixed upon fabrication.

## III. ARTIFICIAL SQUARE SKYRMION ICE

### A. Vertex configuration of artificial square skyrmion ice

In the frustrated square skyrmion crystal, as shown in Fig. 2(c), four holes meet to form a vertex corresponding to the oxygen atoms or pyrochlore tetrahedrons. The captured skyrmions can adopt various configurations that model ''in'' or ''out'' spins. An effective spin vector is assigned to each hole that points toward the end of the hole in which the skyrmion is sitting, so that the state of each hole is defined as spin ''in'' pointing into the vertex if the skyrmion sits close to the vertex and ''out'' otherwise. The square skyrmion ice system can be described by a 16 vertex model which can be categorized into four vertex types according to the skyrmion arrangement as illustrated in Fig. 3. Type I and Type II configurations are ice-rule-obeying states with two skyrmions close to the vertex and two skyrmions away from the vertex. While both Type I and Type II configurations obey the two-spins-in/two-spins-out ice rule, they are energetically split by the square ice geometry. Type I vertices have a twofold degenerate ground state, while Type II vertices possess dipole moments with a fourfold degeneracy. The Type III and Type IV configurations incorporate ice-rule-breaking states. Type III vertices possess a single 'monopole'-like vertex state with a three-spins-in/one-spin-out or three-spins-out/one-spin-in configuration. Type IV vertices form a double 'monopole'-like vertex state with a four-spins-in or four-spins-out configuration. The realization of skyrmion spin ice differs from the nanomagnet system, where north-north and south-south magnetic interactions at a vertex have equal energy. For the skyrmions, interactions between two filled hole ends the vertex energy, whereas two adjacent empty hole ends decrease the vertex energy. Due to the conservation of skyrmion number, creating empty hole ends at one vertex increases the skyrmion load at neighboring vertices. As a result, the ice rules still apply to the skyrmion ice system, but they arise due to collective effects rather than from a local energy minimization.

We consider an ordered 40 × 40 arrangement of vertices with periodic boundary conditions in the *x* and *y* directions containing 3200 blind holes that capture skyrmions. In Fig. 4(a), we illustrate the as-relaxed vertex configuration of the proposed skyrmion ice system. Different vertex types are indicated by different colors depending on the number of skyrmions near each vertex as illustrated at the top of Fig. 4(a). We use the nomenclature $N_0$ for the four-skyrmion-out +2 double monopole vertices; $N_1$ for the one-skyrmion-in/three-skyrmion-out +1 monopole vertices; $N_2$ for the two-skyrmion-in/two-skyrmion-out vertices; $N_3$ for the three-skyrmion-in/one-skyrmion-out −1 monopole vertices; and $N_4$ for the four-skyrmion-in -2 double monopole vertices. The $N_2$ vertices are further subdivided into biased vertices $N_{2\text{-bi}}$ where the two close skyrmions are in adjacent holes and ground state vertices $N_{2\text{-gs}}$ where the two close skyrmions are on opposite sides of the vertex. The as-relaxed ice system is randomly occupied by each of the six vertex types. $N_1$, $N_{2\_bi}$, and $N_3$ vertices are the most common, while $N_0$, $N_{2\_gs}$, and $N_4$ vertices are uncommon.

## B. Spin-torque reconfiguration of skyrmion ice state

To investigate the effect of an external drive, we apply an in-plane spin-polarized electron current $j_s$ (with sign opposite to the electric current $j$) along the positive x-axis. Application of an external spin-polarized current can drive the skyrmions along the conduction electron flow direction through the STT induced dissipative drag force [5,22,23,78,79,89,90]. Skyrmions also experience a non-dissipative Magnus force perpendicular to their velocity due to the underlying emergent electromagnetic field [4,5,19,23,78,80,81,91]. It is possible to counter the current-induced drag force and Magnus force by surrounding the skyrmion with artificial pinning potential barriers produced by blind holes. The skyrmion is driven mostly by the field-like torque from the STT for the in-plane driving case [83]. Under the application of an in-plane current, the skyrmion moves at an angle with respect to the conduction electron flow when $\alpha \neq \beta$ [23,92], where $\alpha$ is the Gilbert damping constant and $\beta$ is the non-adiabatic constant of the spin transfer torque. The movement of the skyrmion away from the intended direction can be attributed to the presence of the Magnus force, which arises due to the coupling between the conduction electrons and the local magnetization [23,78,91]. Under the current-induced drag force and Magnus force, the skyrmions shift within the blind holes to balance the net external force acting on them. Thus, the vertex configuration in the skyrmion ice system can be changed by the application of an external driving current.

For intermediate $j_s$, we mainly observe $N_{2\_bi}$ vertices with a few monopole pairs of $N_1$ and $N_3$ vertices as shown in Fig. 4(b). Each $N_3$ monopole excitation must have a compensating $N_1$ monopole excitation. Such ±1 excitation pairs can be created through a single spin flip, while subsequent spin flips make it possible for the magnetic charges to move some distance away from each other through the lattice. For sufficiently large $j_s$, a biased skyrmion ice configuration containing only $N_{2\_bi}$ vertices forms, as illustrated in Fig. 4(c), with skyrmions located in the bottom ends of the vertically oriented holes and the right ends of the horizontally oriented holes. The variation of $N_i/N$ as a function of the magnitude of the electron current density $j_s$ applied along the positive *x*-axis is plotted in Fig. 5(a). The system starts from the as-relaxed random vertex state shown in Fig. 5(b). The $N_0$, $N_{2\_gs}$, and $N_4$ vertices disappear rapidly when $j_s$ is larger than the threshold density $7 \times 10^{10}$ A/m$^2$. The system attempts to minimize its energy by creating as many $N_{2\_bi}$ vertices as possible with a few monopole pairs of $N_1$ and $N_3$ vertices as shown in Fig. 5(c). As $j_s$ increases past $j_s = 20 \times 10^{10}$ A/m$^2$, the ice system enters the positively biased ice-rule-obeying state illustrated in Fig. 5(d) containing only $N_{2\_bi}$ vertices with $N_{2\_bi}/N = 1$.

Since the skyrmion ice system can be polarized into a biased state by an externally applied current with a particular orientation, it is also possible to create vertex states in which the net sum of the effective spin vectors has a finite value. This permits an effective magnetization hysteresis loop to be constructed under a varying current that is analogous to the hysteretic magnetization versus external magnetic field curves for real spin systems. We define the reduced magnetization *m* as the net sum of the effective spin of each hole [58]:

$$\boldsymbol{m} = N^{-1} \sum_{i=1}^{N} \boldsymbol{s}_i^{eff} \square (\boldsymbol{x} + \boldsymbol{y}) \qquad (3)$$

where $N$ is the number of blind holes, $\boldsymbol{s}_i^{eff}$ is a unit vector defined to point toward the skyrmion-occupied end of the hole, and $\boldsymbol{x}$ and $\boldsymbol{y}$ are the unit vector along the *x*- and *y*-axis, respectively. Figure 5(e) shows a hysteresis loop generated by determining the net local magnetic moment of individual holes while sweeping the applied current. The loop exhibits two-step switching as can be understood by considering the behaviour of the two orthogonal sublattices of the square ice system. The skyrmions trapped in the vertical sublattice can more easily be driven by the current than those trapped in the horizontal sublattice, producing two different coercive current values [59]. At $j_s =$ $30 \times 10^{10}$ A/m$^2$, the overall system obeys the ice rules, forming a biased state with a positively saturated effective magnetization $m = +1$ as illustrated in Fig. 5(f). The skyrmions in the horizontally/vertically oriented holes are pushed to the right/bottom end of each hole. For the downward $m$ curve generated by reducing $j_s$ from $30 \times 10^{10}$ A/m$^2$, the positively saturated biased state persists until $j_s = -6 \times 10^{10}$ A/m$^2$, at which point a sharp step begins, with $m$ reaching 0 as the vertical sublattice switches to form a non-saturated biased state shown in Fig. 5(g). On decreasing $j_s$ towards $-30 \times 10^{10}$ A/m$^2$, $m$ approaches negative saturation sharply at $j_s \approx -20 \times 10^{10}$ A/m$^2$, and the horizontal sublattice switches to form a negatively saturated biased state as shown in Fig. 5(h). The upward $m$ curve behaves like the downward $m$ curve reflected through the $m$ and $j_s$ axis (*i.e.* $m_{\text{downward}}(j_s) = -m_{\text{upward}}(-j_s)$). The sharp steps observed are due to the sub-lattice holes switching. Therefore, our proposed skyrmion ice system captures hysteretic behavior similar to that observed in artificial spin ice systems under applied external magnetic fields.

### C. Ice ground state of square skyrmion ice

The reliable and repeatable acquisition of square ice ground state (GS) order has been a desirable goal [66,93]. Despite theoretical predictions [47], however, a complete GS has not been achieved in nanomagnetic ice systems; instead, only short-range-ordered GS can be reached experimentally by rotating-field demagnetization protocols [43,70]. Only recently, the long-range-ordered ground state has been achieved nearly perfectly over extended regions using a single-shot thermal annealing technique during fabrication [51,94] and a thermally induced melting-freezing protocol is presented to explore experimentally the formation of thermally induced long-range ground-state ordering in a square artificial spin-ice systems of elongated ferromagnetic nanoislands [95–97]. A long-range ground-state magnetic ordering region separated by chains of higher energy vertex configurations has been observed. Our results indicate that the proposed skyrmion system can serve as a model for artificial ice with a random as-relaxed state made up of all vertex configurations rather than an ordered ground state containing only Type I vertices. In order to reliably and repeatedly produce a long-range-ordered square ice GS for the square skyrmion ice systems, we propose a straightforward geometrical protocol. By changing the shape of the blind hole from a symmetrical ellipse to an asymmetrical bullet, the trapped skyrmions tend to occupy the wide end of the holes after relaxation. Hence, it is possible to obtain the ice rule ground state of a square skyrmion ice by arranging these asymmetrical bullet-shaped blind holes in a square lattice with four sub-lattices as shown in Fig. 6(a). Figure 6(b) illustrates that the ice system relaxes into a non-random state filled entirely with Type I vertices in a checkerboard pattern corresponding to the complete square ice GS. The magnetic monopole defects are fully annihilated.

The hysteresis loop of our modified square skyrmion ice system appears in Fig. 6(c). In contrast to the two-step hysteresis loop for the ice system with elliptical blind holes, we find a multi-step switching since the square lattice is effectively composed of four sublattices, labeled $A$ to $D$ in Fig. 6(a). Each lattice has different switching fields depending on the current direction. For a sufficiently large current ($j_s > 64 \times 10^{10}$ A/m$^2$) in the positive $x$-direction, the system enters the positively saturated biased state, $m = +1$, illustrated in Fig. 6(d). The skyrmions in sublattices $A$ and $B$ switch in the negative $y$-direction while the skyrmions in sublattices $C$ and $D$ switch in the positive $x$-direction. Upon reducing $j_s$ to $-14 \times 10^{10}$ A/m$^2$, sublattice $A$ switches in the positive $y$-direction first due to its lower threshold, but since the skyrmions in the other three sublattices have not yet switched, an ordered monopole state forms which consists of an ordered lattice of $N_1$ and $N_3$ vertices as shown in Fig. 6(e). As $j_s$ continuously decreases to $-26 \times 10^{10}$ A/m$^2$, sublattice $B$ switches and the system reaches the non-saturated biased state shown in Fig. 6(f). At $j_s = -36 \times 10^{10}$ A/m$^2$, sublattice $C$ switches and the system approaches an ordered monopole state as shown in Fig. 6(g). Finally, with the switching of sublattice D for $j_s < -64 \times 10^{10}$ A/m$^2$, the system is negatively saturated, $m = -1$, as illustrated in Fig. 6(h). A similar switching behavior occurs when increasing $j_s$ from -70 to +70 × 10$^{10}$ A/m$^2$, and the sublattices switch continuously in the order B, A, D, and C, as illustrated in Figs. 6(h)-6(k).

### D. Boundary effects

We have presented the vertex configuration of square skyrmion ice including an ordered 40 × 40 arrangement of vertices with periodic boundary conditions. It is important to understand whether the boundaries affect the response of the system as experimental systems have no periodic boundary conditions. We therefore test whether a small system can still exhibit the artificial ice behavior observed for larger arrays with either ordered or disordered configurations. Figure 7 shows the vertex configuration of a finite square skyrmion ice system containing 60 blind holes with open boundary conditions. This finite square skyrmion ice system exhibits similar vertex configurations as those observed in the larger system. The as-relaxed vertex configuration exhibits an ice-rule-obeying ground state as illustrated in Fig. 7(a). With the application of an external driving current, the vertex configuration can be driven into an ice-rule-breaking disordered state, an ice-rule-breaking monopole state, and an ice-rule-obeying biased state as shown in Figs. 7(b)-7(d), respectively. In addition, the artificial ice behavior observed for the square skyrmion ice investigated here also appears for honeycomb skyrmion ice with a hexagonal arrangement of blind holes.

## IV. ARTIFICIAL HONEYCOMB SKYRMION ICE

We have studied the geometrical frustration of a square skyrmion crystal system, and this can also be extended to different types of skyrmion crystals by arranging the blind holes in diverse crystal geometries. As an example, we also investigated an artificial honeycomb skyrmion ice system with a honeycomb arrangement of ellipse-shaped blind holes (see Fig. 8(a)). Here, each vertex consists of three equivalent effective spin vectors enclosing an angle of 120°. For each vertex there are a total of 8 possible configurations, which can be categorized into two types according to the arrangement of skyrmions near each vertex as shown in Fig. 8(b). In analogy to the square skyrmion ice system, the six Type I spin configurations obey the two-spins-in-one-spin-out/two-spins-out-one-spin-in spin ice rule for the honeycomb lattice with two skyrmions close to the vertex (two spins point in) and one skyrmion close to the vertex (one points out), or vice versa. The remaining two Type II spin configurations possess a 'monopole'-like vertex state with all three spins pointing either in or out, violating the spin ice rule. Figure 8(c) shows the as-relaxed vertex configuration of the honeycomb skyrmion ice system. Different vertex types are indicated by different colors depending on the number of skyrmions near each vertex as illustrated at the top of Fig. 8(c). We use the nomenclature $N_0$ for the three-skyrmion-out monopole vertices; $N_{1-gs}$ for the one-skyrmion-in/two-skyrmion-out ground state vertices; $N_{2-gs}$ for the two-skyrmion-in/one-skyrmion-out ground state vertices; and $N_3$ for the three-skyrmion-in monopole vertices. The as-relaxed ice system is randomly occupied by each of the four vertex types. It is worth noting that the honeycomb skyrmion ice system can be driven into a complete long-range ordered ground state containing only Type I vertices as well as a complete monopole state containing only Type II vertices by applying an in-plane spin-polarized electron current to shift the positions of skyrmions captured in the holes.

## V. DISCUSSION AND CONCLUSIONS

We have demonstrated the artificial skyrmion spin ice concept using the Néel type skyrmion as shown in FIG. 1(c), while the Bloch type skyrmions observed in natural materials can also be used to investigate the ice configuration [98]. Additionally, the experimentally demonstrated artificial skyrmion systems [28–30] are potential candidates for adjusting the skyrmion position through nanostructuring and driving protocols. The possibility of changing the spin ice configuration of the proposed artificial skyrmion spin ice system with the application of spin-polarized current through the spin transfer torque effect has been demonstrated. In principle the spin ice geometries could be produced using other skyrmion systems, provided that the skymions have an effective repulsive interaction that can give rise to the ice rules. Recently, thermal gradients [99–102], electric field [103], and microwave magnetic field [104] have been used to drive skyrmions into motion. Hence, the ice state reconfiguration of the skyrmion ices can also be alternatively realized by thermal gradients, electric field, and microwave magnetic field.

In summary, we demonstrated a frustrated skyrmion crystal, in which skyrmions located on the non-frustrated square/honeycomb lattice show an unexpectedly exotic ice phase. The existence of such frustrated skyrmion states is an example of geometrical frustration emerging from competing interactions between structural confinement and external drives. The system reveals a rich phase behavior when skyrmion-current interactions compete with structural confinement strength. In contrast to lattices of interacting nanoscale particles/islands such as artificial spin ice [43,44], or colloids [71,72,105], the spin-torque reconfigurability of the proposed skyrmion ices makes them compatible with standard spintronic devices such as magnetic tunnel junctions and thus straightforwardly integrable to the existing semiconductor manufacturing processes. Therefore, the artificial skyrmion ice system represents a versatile model to investigate geometrically frustrated states and provides an entirely new paradigm to investigate the effect of vertex configurations on ordered and degenerate ground states, avalanche dynamics, return point memory, and the hopping and annihilating of monopole-like defects.

ACKNOWLEDGMENTS

This work was supported by the Singapore National Research Foundation, Prime Minister's Office, under a Competitive Research Programme (Non-volatile Magnetic Logic and Memory Integrated Circuit Devices, NRF-CRP9-2011-01). W.S.L. is a member of the Singapore Spintronics Consortium (SG-SPIN).

**References**

[1]  U. K. Rössler, A. N. Bogdanov, and C. Pfleiderer, Nature **442**, 797 (2006).
[2]  I. Dzyaloshinsky, J. Phys. Chem. Solids **4**, 241 (1958).
[3]  T. Moriya, Phys. Rev. **120**, 91 (1960).
[4]  S. Mühlbauer, B. Binz, F. Jonietz, C. Pfleiderer, A. Rosch, A. Neubauer, R. Georgii, and P. Böni, Science **323**, 915 (2009).
[5]  X. Z. Yu, Y. Onose, N. Kanazawa, J. H. Park, J. H. Han, Y. Matsui, N. Nagaosa, and Y. Tokura, Nature **465**, 901 (2010).
[6]  X. Yu, J. P. DeGrave, Y. Hara, T. Hara, S. Jin, and Y. Tokura, Nano Lett. **13**, 3755 (2013).
[7]  D. Liang, J. P. DeGrave, M. J. Stolt, Y. Tokura, and S. Jin, Nat. Commun. **6**, 8217 (2015).
[8]  H. Du, D. Liang, C. Jin, L. Kong, M. J. Stolt, W. Ning, J. Yang, Y. Xing, J. Wang, R. Che, J. Zang, S. Jin, Y. Zhang, and M. Tian, Nat. Commun. **6**, 7637 (2015).
[9]  H. Du, J. P. DeGrave, F. Xue, D. Liang, W. Ning, J. Yang, M. Tian, Y. Zhang, and S. Jin, Nano Lett. **14**, 2026 (2014).
[10]  A. Mehlin, F. Xue, D. Liang, H. F. Du, M. J. Stolt, S. Jin, M. L. Tian, and M. Poggio, Nano Lett. **15**, 4839 (2015).
[11]  S. Heinze, K. von Bergmann, M. Menzel, J. Brede, A. Kubetzka, R. Wiesendanger, G. Bihlmayer, and S. Blügel, Nat. Phys. **7**, 713 (2011).
[12]  N. Romming, C. Hanneken, M. Menzel, J. E. Bickel, B. Wolter, K. von Bergmann, A. Kubetzka, and R. Wiesendanger, Science **341**, 636 (2013).
[13]  F. Büttner, C. Moutafis, M. Schneider, B. Krüger, C. M. Günther, J. Geilhufe, C. v. K. Schmising, J. Mohanty, B. Pfau, S. Schaffert, A. Bisig, M. Foerster, T. Schulz, C. A. F. Vaz, J. H. Franken, H. J. M. Swagten, M. Kläui, and S. Eisebitt, Nat. Phys. **11**, 225 (2015).
[14]  W. Jiang, P. Upadhyaya, W. Zhang, G. Yu, M. B. Jungfleisch, F. Y. Fradin, J. E. Pearson, Y. Tserkovnyak, K. L. Wang, O. Heinonen, S. G. E. te Velthuis, and A. Hoffmann, Science **349**, 283 (2015).
[15]  S. Woo, K. Litzius, B. Krüger, M.-Y. Im, L. Caretta, K. Richter, M. Mann, A. Krone, R. M. Reeve, M. Weigand, P. Agrawal, I. Lemesh, M.-A. Mawass, P. Fischer, M. Kläui, and G. S. D. Beach, Nat. Mater. **15**, 501 (2016).
[16]  C. Moreau-Luchaire, C. Moutafis, N. Reyren, J. Sampaio, C. A. F. Vaz, N. Van Horne, K. Bouzehouane, K. Garcia, C. Deranlot, P. Warnicke, P. Wohlhüter, J.-M. George, M. Weigand, J. Raabe, V. Cros, and A. Fert, Nat. Nanotechnol. **11**, 444 (2016).
[17]  O. Boulle, J. Vogel, H. Yang, S. Pizzini, D. de Souza Chaves, A. Locatelli, T. O. Menteş, A. Sala, L. D. Buda-Prejbeanu, O. Klein, M. Belmeguenai, Y. Roussigné, A. Stashkevich,

S. M. Chérif, L. Aballe, M. Foerster, M. Chshiev, S. Auffret, I. M. Miron, and G. Gaudin, Nat. Nanotechnol. **11**, 449 (2016).

[18] A. Soumyanarayanan, M. Raju, A. L. G. Oyarce, A. K. C. Tan, M.-Y. Im, A. P. Petrovic, P. Ho, K. H. Khoo, M. Tran, C. K. Gan, F. Ernult, and C. Panagopoulos, arXiv:1606.06034 (2016).
[19] N. Nagaosa and Y. Tokura, Nat. Nanotechnol. **8**, 899 (2013).
[20] J. Sampaio, V. Cros, S. Rohart, A. Thiaville, and A. Fert, Nat. Nanotechnol. **8**, 839 (2013).
[21] A. Fert, V. Cros, and J. Sampaio, Nat. Nanotechnol. **8**, 152 (2013).
[22] F. Jonietz, S. Mühlbauer, C. Pfleiderer, A. Neubauer, W. Münzer, A. Bauer, T. Adams, R. Georgii, P. Böni, R. A. Duine, K. Everschor, M. Garst, and A. Rosch, Science **330**, 1648 (2010).
[23] T. Schulz, R. Ritz, A. Bauer, M. Halder, M. Wagner, C. Franz, C. Pfleiderer, K. Everschor, M. Garst, and A. Rosch, Nat. Phys. **8**, 301 (2012).
[24] X. Zhang, M. Ezawa, and Y. Zhou, Sci. Rep. **5**, 9400 (2015).
[25] F. Ma, Y. Zhou, H. B. Braun, and W. S. Lew, Nano Lett. **15**, 4029 (2015).
[26] H.-B. Braun, Adv. Phys. **61**, 1 (2012).
[27] L. Sun, R. X. Cao, B. F. Miao, Z. Feng, B. You, D. Wu, W. Zhang, A. Hu, and H. F. Ding, Phys. Rev. Lett. **110**, 167201 (2013).
[28] B. F. Miao, L. Sun, Y. W. Wu, X. D. Tao, X. Xiong, Y. Wen, R. X. Cao, P. Wang, D. Wu, Q. F. Zhan, B. You, J. Du, R. W. Li, and H. F. Ding, Phys. Rev. B **90**, 174411 (2014).
[29] J. Li, A. Tan, K. W. Moon, A. Doran, M. A. Marcus, A. T. Young, E. Arenholz, S. Ma, R. F. Yang, C. Hwang, and Z. Q. Qiu, Nat. Commun. **5**, 4704 (2014).
[30] D. A. Gilbert, B. B. Maranville, A. L. Balk, B. J. Kirby, P. Fischer, D. T. Pierce, J. Unguris, J. A. Borchers, and K. Liu, Nat. Commun. **6**, 8462 (2015).
[31] Y. Y. Dai, H. Wang, P. Tao, T. Yang, W. J. Ren, and Z. D. Zhang, Phys. Rev. B **88**, 054403 (2013).
[32] M. V. Sapozhnikov and O. L. Ermolaeva, Phys. Rev. B **91**, 024418 (2015).
[33] A. I. Marchenko and V. N. Krivoruchko, J. Magn. Magn. Mater. **377**, 153 (2015).
[34] M. V. Sapozhnikov, J. Magn. Magn. Mater. **396**, 338 (2015).
[35] L. Pauling, J. Am. Chem. Soc. **57**, 2680 (1935).
[36] P. W. Anderson, Phys. Rev. **102**, 1008 (1956).
[37] M. J. Harris, S. T. Bramwell, D. F. McMorrow, T. Zeiske, K. W. Godfrey, and K. W. Godfrey, Phys. Rev. Lett. **79**, 2554 (1997).
[38] M. J. Harris, S. T. Bramwell, P. C. W. Holdsworth, and J. D. M. Champion, Phys. Rev. Lett. **81**, 4496 (1998).
[39] A. P. Ramirez, A. Hayashi, R. J. Cava, R. Siddharthan, and B. S. Shastry, Nature **399**, 333 (1999).
[40] S. T. Bramwell, Science **294**, 1495 (2001).
[41] C. Castelnovo, R. Moessner, and S. L. Sondhi, Nature **451**, 42 (2008).
[42] I. Gilbert, C. Nisoli, and P. Schiffer, Phys. Today **69**, 54 (2016).
[43] R. F. Wang, C. Nisoli, R. S. Freitas, J. Li, W. McConville, B. J. Cooley, M. S. Lund, N. Samarth, C. Leighton, V. H. Crespi, and P. Schiffer, Nature **439**, 303 (2006).
[44] C. Nisoli, R. Moessner, and P. Schiffer, Rev. Mod. Phys. **85**, 1473 (2013).
[45] L. J. Heyderman and R. L. Stamps, J. Phys. Condens. Matter **25**, 363201 (2013).
[46] J. Cumings, L. J. Heyderman, C. H. Marrows, and R. L. Stamps, New J. Phys. **16**, 075016 (2014).

[47] G. Möller, R. Moessner, R. Moessner, R. Moessner, and R. Moessner, Phys. Rev. Lett. **96**, 237202 (2006).
[48] C. Nisoli, J. Li, X. Ke, D. Garand, P. Schiffer, and V. H. Crespi, Phys. Rev. Lett. **105**, 047205 (2010).
[49] P. E. Lammert, X. Ke, J. Li, C. Nisoli, D. M. Garand, V. H. Crespi, P. Schiffer, P. Schiffer, and P. Schiffer, Nat. Phys. **6**, 786 (2010).
[50] C. Nisoli, R. Wang, J. Li, W. F. McConville, P. E. Lammert, P. Schiffer, and V. H. Crespi, Phys. Rev. Lett. **98**, 217203 (2007).
[51] J. P. Morgan, A. Stein, S. Langridge, and C. H. Marrows, Nat. Phys. **7**, 75 (2011).
[52] E. Mengotti, L. J. Heyderman, A. F. Rodríguez, F. Nolting, R. V. Hügli, and H.-B. Braun, Nat. Phys. **7**, 68 (2011).
[53] S. Ladak, D. E. Read, G. K. Perkins, L. F. Cohen, and W. R. Branford, Nat. Phys. **6**, 359 (2010).
[54] N. Rougemaille, F. Montaigne, B. Canals, A. Duluard, D. Lacour, M. Hehn, R. Belkhou, O. Fruchart, S. El Moussaoui, A. Bendounan, and F. Maccherozzi, Phys. Rev. Lett. **106**, 057209 (2011).
[55] S. A. Daunheimer, O. Petrova, O. Tchernyshyov, and J. Cumings, Phys. Rev. Lett. **107**, 167201 (2011).
[56] Z. Budrikis, J. P. Morgan, J. Akerman, A. Stein, P. Politi, S. Langridge, C. H. Marrows, and R. L. Stamps, Phys. Rev. Lett. **109**, 037203 (2012).
[57] W. R. Branford, S. Ladak, D. E. Read, K. Zeissler, and L. F. Cohen, Science **335**, 1597 (2012).
[58] A. Libál, C. Reichhardt, and C. J. O. Reichhardt, Phys. Rev. Lett. **97**, 228302 (2006).
[59] C. J. Olson Reichhardt, A. Libál, and C. Reichhardt, New J. Phys. **14**, 025006 (2012).
[60] A. Libál, C. J. O. Reichhardt, and C. J. O. Reichhardt, Phys. Rev. Lett. **102**, 237004 (2009).
[61] M. L. Latimer, G. R. Berdiyorov, Z. L. Xiao, F. M. Peeters, and W. K. Kwok, Phys. Rev. Lett. **111**, 067001 (2013).
[62] J. Trastoy, M. Malnou, C. Ulysse, R. Bernard, N. Bergeal, G. Faini, J. Lesueur, J. Briatico, and J. E. Villegas, Nat. Nanotechnol. **9**, 710 (2014).
[63] Z. Budrikis, P. Politi, and R. L. Stamps, Phys. Rev. Lett. **105**, 017201 (2010).
[64] P. Mellado, O. Petrova, Y. Shen, and O. Tchernyshyov, Phys. Rev. Lett. **105**, 187206 (2010).
[65] Z. Budrikis, P. Politi, and R. L. Stamps, Phys. Rev. Lett. **107**, 217204 (2011).
[66] X. Ke, J. Li, C. Nisoli, P. E. Lammert, W. McConville, R. F. Wang, V. H. Crespi, P. Schiffer, and P. Schiffer, Phys. Rev. Lett. **101**, 037205 (2008).
[67] A. Farhan, P. M. Derlet, A. Kleibert, A. Balan, R. V. Chopdekar, M. Wyss, L. Anghinolfi, F. Nolting, and L. J. Heyderman, Nat. Phys. **9**, 375 (2013).
[68] E. Mengotti, L. J. Heyderman, A. F. Rodriguez, A. Bisig, L. Le Guyader, F. Nolting, and H. B. Braun, Phys. Rev. B **78**, 144402 (2008).
[69] Y. Y. Qi, T. Brintlinger, and J. Cumings, Phys. Rev. B **77**, 094418 (2008).
[70] J. P. Morgan, A. Bellew, A. Stein, S. Langridge, and C. H. Marrows, Front. Phys. **1**, 28 (2013).
[71] P. Tierno, Phys. Rev. Lett. **116**, 038303 (2016).
[72] A. Ortiz-Ambriz and P. Tierno, Nat. Commun. **7**, 10575 (2016).
[73] C. Moutafis, S. Komineas, and J. A. C. Bland, Phys. Rev. B **79**, 224429 (2009).
[74] X. Zhao, C. Jin, C. Wang, H. Du, J. Zang, M. Tian, R. Che, and Y. Zhang, Proc. Natl. Acad.

Sci. **113**, 4918 (2016).
[75] A. A. Thiele, Phys. Rev. Lett. **30**, 230 (1973).
[76] K. Everschor, M. Garst, R. A. Duine, and A. Rosch, Phys. Rev. B **84**, 064401 (2011).
[77] K. Everschor, M. Garst, B. Binz, F. Jonietz, S. Mühlbauer, C. Pfleiderer, and A. Rosch, Phys. Rev. B **86**, 054432 (2012).
[78] J. Iwasaki, M. Mochizuki, and N. Nagaosa, Nat. Commun. **4**, 1463 (2013).
[79] J. Zang, M. Mostovoy, J. H. Han, and N. Nagaosa, Phys. Rev. Lett. **107**, 136804 (2011).
[80] S.-Z. Lin, C. Reichhardt, C. D. Batista, and A. Saxena, Phys. Rev. B **87**, 214419 (2013).
[81] C. Reichhardt, D. Ray, and C. J. O. Reichhardt, Phys. Rev. Lett. **114**, 217202 (2015).
[82] S.-Z. Lin, C. Reichhardt, C. D. Batista, and A. Saxena, J. Appl. Phys. **115**, 17D109 (2014).
[83] A. Thiaville, Y. Nakatani, J. Miltat, and Y. Suzuki, Europhys. Lett. **69**, 990 (2005).
[84] A. Vansteenkiste, J. Leliaert, M. Dvornik, M. Helsen, F. Garcia-Sanchez, and B. Van Waeyenberge, AIP Adv. **4**, 107133 (2014).
[85] S. Rohart and A. Thiaville, Phys. Rev. B **88**, 184422 (2013).
[86] A. Thiaville, S. Rohart, É. Jué, V. Cros, and A. Fert, EPL **100**, 57002 (2012).
[87] F. Ma and Y. Zhou, RSC Adv. **4**, 46454 (2014).
[88] S. Zhang and Z. Li, Phys. Rev. Lett. **93**, 127204 (2004).
[89] X. Z. Yu, N. Kanazawa, W. Z. Zhang, T. Nagai, T. Hara, K. Kimoto, Y. Matsui, Y. Onose, and Y. Tokura, Nat. Commun. **3**, 988 (2012).
[90] S.-Z. Lin, C. Reichhardt, C. D. Batista, and A. Saxena, Phys. Rev. Lett. **110**, 207202 (2013).
[91] J. Iwasaki, M. Mochizuki, and N. Nagaosa, Nat. Nanotechnol. **8**, 742 (2013).
[92] A. Brataas, A. D. Kent, and H. Ohno, Nat. Mater. **11**, 372 (2012).
[93] Y. Y. Qi, T. Brintlinger, J. Cumings, J. Cumings, J. Cumings, and J. Cumings, Phys. Rev. B **77**, 094418 (2008).
[94] Jason P. Morgan, J. Akerman, A. Stein, C. Phatak, R. M. L. Evans, S. Langridge, and C. H. Marrows, Phys. Rev. B **87**, 024405 (2013).
[95] A. Farhan, P. M. Derlet, A. Kleibert, A. Balan, R. V. Chopdekar, M. Wyss, J. Perron, A. Scholl, F. Nolting, and L. J. Heyderman, Phys. Rev. Lett. **111**, 057204 (2013).
[96] J. M. Porro, A. Bedoya-Pinto, A. Berger, and P. Vavassori, New J. Phys. **15**, 055012 (2013).
[97] S. Zhang, I. Gilbert, C. Nisoli, G.-W. Chern, M. J. Erickson, L. O'Brien, C. Leighton, P. E. Lammert, V. H. Crespi, and P. Schiffer, Nature **500**, 553 (2013).
[98] I. Kézsmárki, S. Bordács, P. Milde, E. Neuber, L. M. Eng, J. S. White, H. M. Rønnow, C. D. Dewhurst, M. Mochizuki, K. Yanai, H. Nakamura, D. Ehlers, V. Tsurkan, and A. Loidl, Nat. Mater. **14**, 1116 (2015).
[99] L. Kong and J. Zang, Phys. Rev. Lett. **111**, 067203 (2013).
[100] S.-Z. Lin, C. D. Batista, C. Reichhardt, and A. Saxena, Phys. Rev. Lett. **112**, 187203 (2014).
[101] A. A. Kovalev, Phys. Rev. B **89**, 241101 (2014).
[102] S. K. Kim, O. Tchernyshyov, and Y. Tserkovnyak, Phys. Rev. B **92**, 020402 (2015).
[103] P. Upadhyaya, G. Yu, P. K. Amiri, and K. L. Wang, Phys. Rev. B **92**, 134411 (2015).
[104] W. Wang, M. Beg, B. Zhang, W. Kuch, and H. Fangohr, Phys. Rev. B **92**, 020403 (2015).
[105] Y. Han, Y. Shokef, A. M. Alsayed, P. Yunker, T. C. Lubensky, and A. G. Yodh, Nature **456**, 898 (2008).

**Figure Captions**

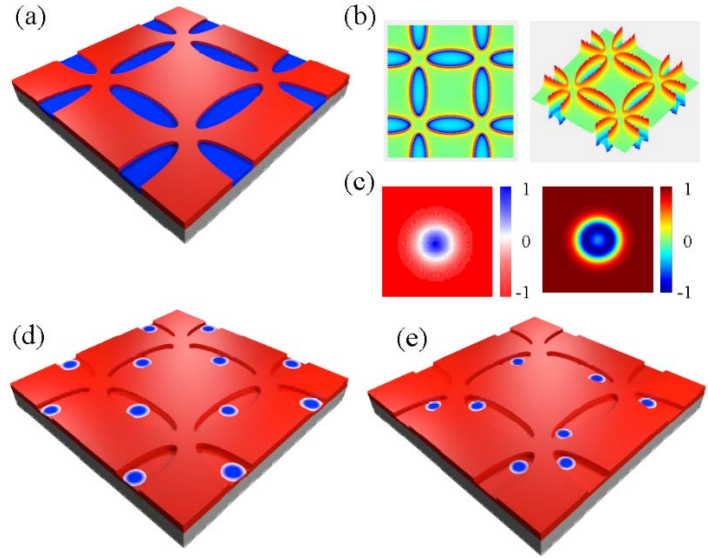

FIG. 1. (Color online) Schematic of proposed artificial square skyrmion ice system. (a) Square array of ellipse-shaped blind holes with the up/down magnetization configuration outside/inside the holes. (b) Calculated perpendicular *z*-component of the stray field in the center of the continuous film generated by the square arrangement of blind holes. (c) Left: The spin configuration of a single magnetic skyrmion. The positive (negative) *z* component of the magnetization is represented by red (blue), whereas white indicates in-plane spin orientation. Right: The topological density distribution of a static skyrmion where the colour-scale is normalized from -1 to +1. Self-assembled skyrmions confined in holes: (d) non-frustrated and (e) frustrated skyrmion crystal.

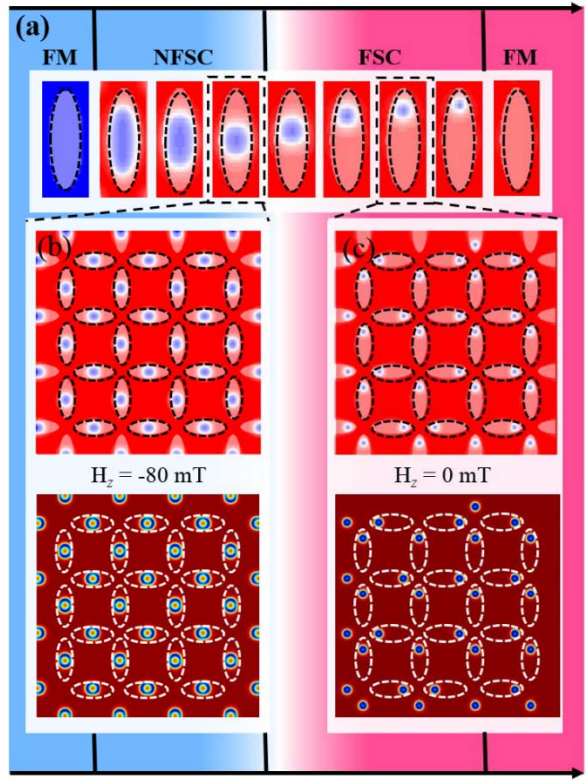

FIG. 2. (Color online) Phase diagram of the magnetization configuration in the magnetic thin film with an array of ellipse-shaped blind holes. FM, NFSC, and FSC denote ferromagnetic, non-frustrated skyrmion crystal, and frustrated skyrmion crystal phases, respectively. The boundary of the hole is indicated by dashed lines. (a) The size and position of a skyrmion captured in a blind hole as a function of the strength of the magnetic field $H_z$. Arrangement of skyrmions assembled in the square array of elliptical potential wells and the corresponding topological density distribution of captured skyrmions in the wells with the boundary indicated by dashed lines: (b) non-frustrated skyrmion crystal phase at $H_z = -80$ mT and (c) frustrated skyrmion crystal phase at $H_z = 0$ mT.

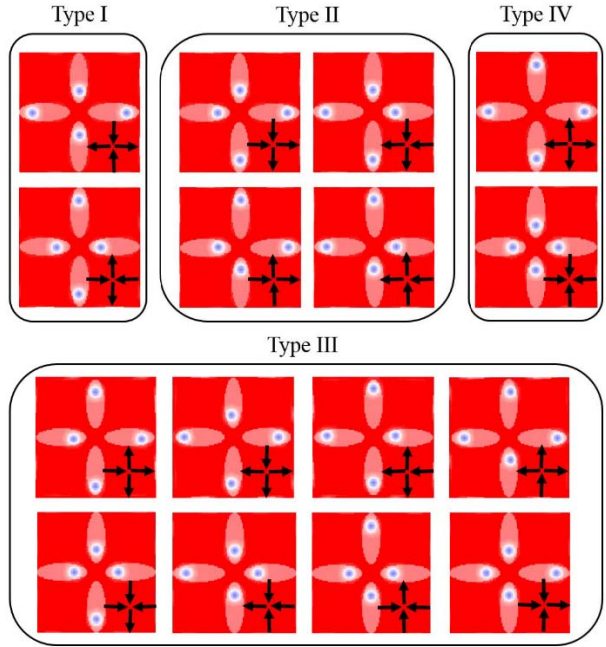

FIG. 3. (Color online) Schematic illustrations of all 16 possible vertex configurations for the square skyrmion ice system categorized into four types according to the arrangement of skyrmions near each vertex. The Type I and Type II configurations obey the spin-ice rule, and the Type I and Type II configurations violate the spin-ice rule. The corresponding moment configurations for each of the vertex types are indicated by solid arrows as insets.

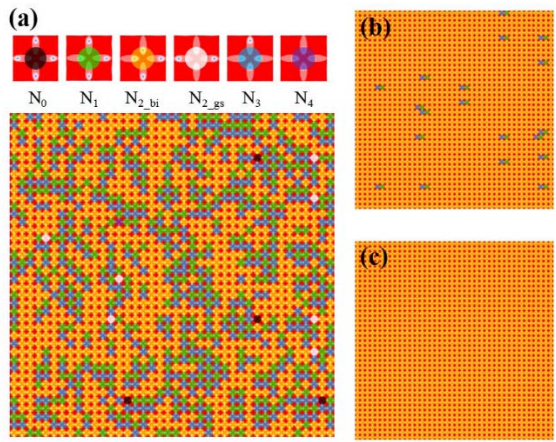

FIG. 4. (Color online) Vertex configuration of the square skyrmion ice system. (a) as-relaxed

random state with all of the six vertex types present, (b) non-saturated biased state containing $N_1$ and $N_3$ monopole pairs, and (c) saturated biased state with only $N_{2\text{-}bi}$ vertices. Vertices are colored depending on how many skyrmions are near each vertex as illustrated at the top of the figure: $N_0$ (gray), $N_1$ (green), $N_{2\text{-}bi}$ (yellow), $N_{2\text{-}gs}$ (white), $N_3$ (blue), and $N_4$ (purple).

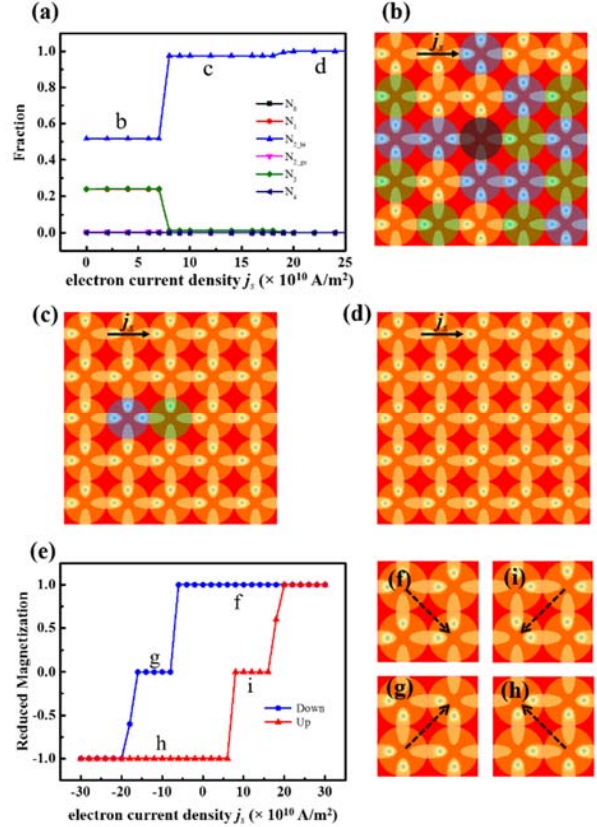

FIG. 5. (Color online) Fraction of vertices and hysteresis loop of reduced magnetization. (a) The fraction of vertices $N_i$ for i = 0 - 4 as a function of the electron density. Vertex configuration of a small portion of the skyrmion ice system at various electron densities as labeled in (a): (b) as-relaxed random state; (c) non-saturated biased state; and (d) positively saturated biased ice rule state. The solid arrows in (b)-(d) indicate the direction of electron density. (e) Hysteresis loop of the reduced magnetization $m$ versus electron current density $j_s$ between $\pm 30 \times 10^{10}$ A/m$^2$ applied along the $x$-axis. Vertex configuration of a small portion of the skyrmion ice system at various electron densities as labeled in (e): (f) positively saturated biased ice rule state; (g) non-saturated biased ice rule state; (h) negatively saturated biased ice rule state; and (i) non-saturated biased ice rule state. Dashed arrows indicate the dipole moment.

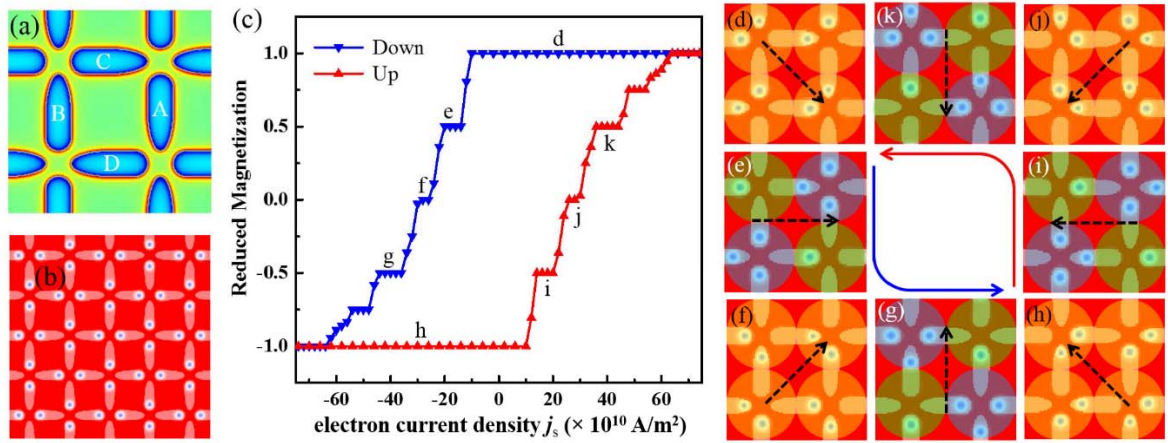

FIG. 6. (Color online) Artificial square skyrmion ice with bullet-shaped blind hole. (a) The calculated stray field of a square array of bullet-shaped blind holes. (b) The ice-rule-obeying ground state configuration filled entirely with Type I vertices. (c) Hysteresis loop of the reduced magnetization $m$ vs electron current density $j_s$ applied along the $x$-axis. Vertex configuration of the skyrmion ice system at various electron current densities as labeled in (c): (d) positively saturated biased ice rule state; (e) monopole state consisting of an ordered lattice of $N_1$ and $N_3$ vertices; (f) non-saturated biased ice rule state; (g) monopole state; (h) negatively saturated biased ice rule state; (i) monopole state; (j) non-saturated biased ice rule state; and (k) monopole state. Dashed arrows indicate the dipole moment.

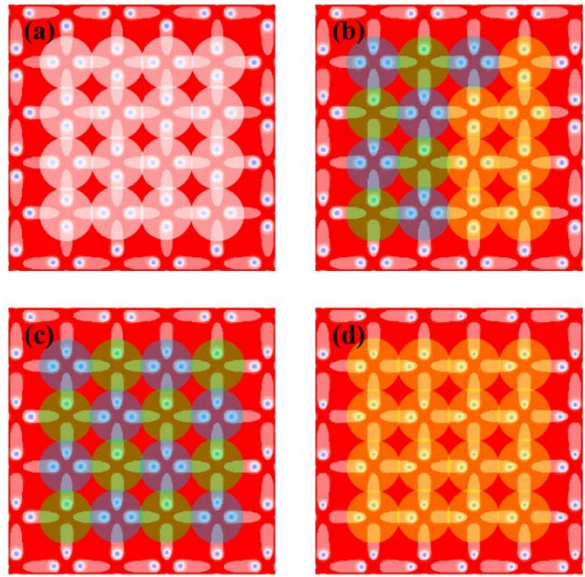

FIG. 7. (Color online) Vertex configuration in square skyrmion ice with open boundary condition.

(a) ice-rule-obeying ground state configuration filled entirely with Type I vertices; (b) disordered state; (c) monopole state consisting of an ordered lattice of $N_1$ and $N_3$ vertices; and (d) biased ice rule state.

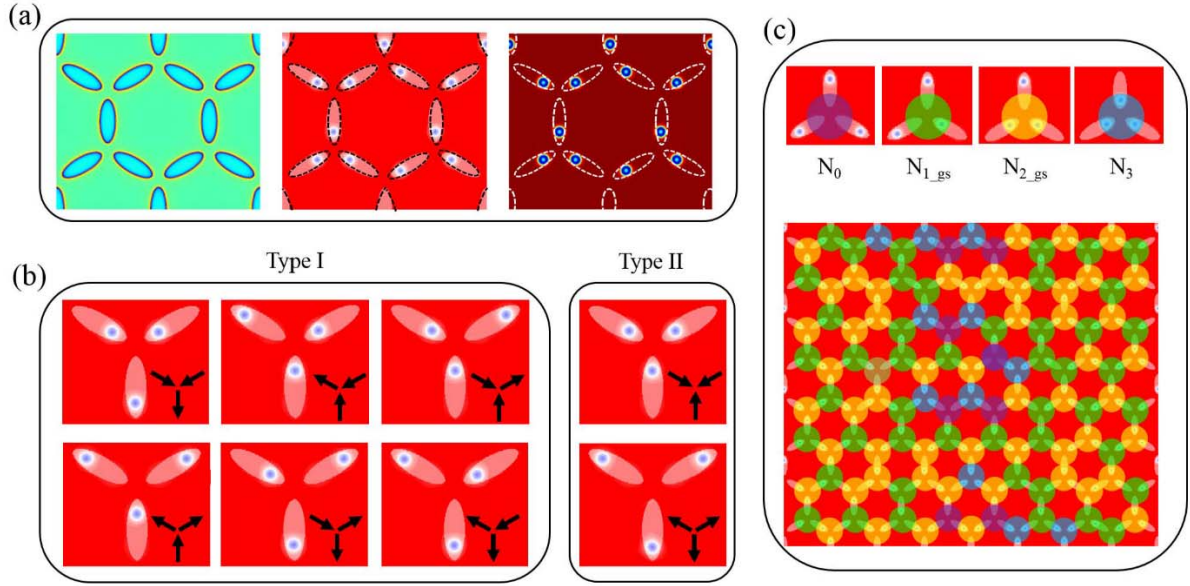

FIG. 8. (Color online) Schematic and vertex configuration of artificial honeycomb skyrmion ice system. (a) Left: Calculated perpendicular $z$-component of the stray field in the center of the continuous film generated by the honeycomb arrangement of blind holes. Middle: Schematic diagram of an artificial honeycomb ice system consisting of a honeycomb array of elliptical potential wells that each capture one skyrmion. Right: The corresponding topological density distribution of captured skyrmions with the potential boundary indicated by dashed lines. (b) Schematic illustrations of all 8 possible vertex configurations for the honeycomb skyrmion ice system categorized into two types according to the arrangement of skyrmions near each vertex. The Type I configuration obeys the spin-ice rule, and the Type II configuration violates the spin-ice rule. The corresponding moment configurations for each of the vertex types are indicated by solid arrows as insets. (c) As-relaxed random vertex configuration of the honeycomb skyrmion ice system with all of the four vertex types present. Vertices are colored depending on how many skyrmions are near each vertex as illustrated at the top of the figure: $N_0$ (purple), $N_{1\_gs}$ (green), $N_{2\text{-}gs}$ (yellow), and $N_3$ (blue).